\documentclass{elsarticle}
\usepackage{amsmath}

\newcommand{\ep}{\varepsilon}

\newcommand{\be}{\begin{equation}}
\newcommand{\ee}{\end{equation}}
\newcommand{\bea}{\begin{eqnarray}}
\newcommand{\eea}{\end{eqnarray}}

\def\re#1{(\ref{#1})}   

\begin{document}

\title{Generic stability in dissipative generalized mechanics}
\author{P. V\'an$^{1,2}$ }
\address{$^1$Dept. of Theoretical Physics, Wigner RCP, RMKI, \\  H-1525 Budapest, P.O.Box 49, Hungary; 
and  {$^2$Dept. of Energy Engineering, Budapest Univ. of Technology and Economics},\\
  H-1111, Budapest, Bertalan Lajos u. 4-6,  Hungary}

\date{\today}

\begin{abstract}
A theory of dissipative generalized continuum mechanics is presented in the framework of weakly nonlocal non-equilibrium thermodynamics. The evolution equation of microdeformation is obtained by thermodynamic principles. Conditions of generic stability, the linear asymptotic stability of homogeneous equilibrium, is derived in a simple but representative case.
\end{abstract}

\maketitle

\section[intro]{Introduction}

There is a kind of stability that is among our basic and minimal physical expectations in continua. Explicitely or implicitely we think, that  small perturbations of a homogeneous thermodynamic equilibrium state decay. Small excitations of arbitrary form are damped and this damping is expected as a result of material properties, otherwise the continuum, as a material, cannot exist, it destroys itself. Theories should reflect this property, or they are wrong models of a natural system. A relatively simple necessary condition of the above mentioned expectation for a specific continuum model is \textit{ generic stability}: the linear asymptotic stability of homogeneous solutions of the evolution equations. If the conditions of generic stability do not require anything beyond thermodynamic reasons, i.e. by the concavity of the entropy and by the nonnegativity of the entropy production, then one can be 
sure that the parameters of the model family represent material properties, the mentioned basic stability expectation is fulfilled. Generic stability may be a benchmark on the viability of theories of dissipation. The more the conditions of generic stability are connected to thermodynamic principles, to concavity of entropy density and increasing entropy, the higher can be our hope that our dissipative theory is a viable model of a natural system.  

The concept of generic stability was introduced for special relativistic dissipative fluids by Hiscock and Lindblom \cite{HisLin85a}. There generic stability distinguishes between thermodynamically consistent looking theories with increasing entropy production. There are violently unstable ones \cite{HisLin85a}, conditionally stable ones \cite{HisLin83a} and there are those where generic stability does not require anything beyond basic thermodynamics requirements \cite{VanBir12a}, and in this sense they are similar to the nonrelativistic, dissipative Navier-Stokes fluids \cite{Van09a}. 

Thermodynamics and stability of some simple equilibria are considered closely related also in solids. Sometimes the expectations are high, failures are interpreted as fatal and any relation is denied \cite{MulWei12a}. On the other hand, new researches about the fundamental connection of the second law and stability of thermodynamic equilibrium \cite{Mat05b} change the failures to challenges that may lead to new approaches to old problems \cite{BirVan11a}. 

In generalized mechanics dissipation looks like a secondary aspect.  Mindlin introduced a variational principle \cite{Min64a}, Eringen and Suhubi used a balance of micromomentum with  averaging \cite{EriSuh64a} and Germain suggested the principle of virtual power \cite{Ger73a} to obtain new evolution equations. None of them was interested in dissipation. The situation did not change much since then, the search for the proper interpretation and nondissipative material parameters suppressed the need of a dissipative generalization \cite{Eri99b}. However, recent research revealed a fundamental connection of thermodynamics and generalized mechanics. From a thermodynamic point of view the microdeformation can be considered as an internal variable and the evolution equations of Mindlin can be obtained by introducing dual tensorial internal variables \cite{BerEta11a}. 

In this paper we obtain the Mindlin-Eringen-Suhubi theory of microdeformation by weakly nonlocal non-equilibrium thermodynamics. Hence we arrive at a dissipative theory from the very beginning. Then we investigate the generic stability of the obtained dissipative theory in a reasonably simple, but representative specific case. It turns out that pure thermodynamics conditions are not enough, instabilities may arise.

\section{Mindlin theory of microdeformation}

The fundamental kinematic quantities of generalized mechanics are the \textit{displacement} $u^i = x^i-X^i$ and the \textit{micro-displacement} $u'^i = x'^i-X'^i$, where $x^i$ and $x'^i$ are spatial $X^i$ and $X'^i$ are material position vectors. Then one introduces the strain as the symmetric part of the displacement gradient $\ep^{ij} =\partial^{(j}u^{i)} =\frac{1}{2}(\partial^i u^j +\partial^j u^i)$, and microdeformation as the gradient of the microdisplacement $\psi^{ij}=\partial^j u'^i$. With displacements and Cartesian coordinates the treatment is frame and coordinate dependent from the beginning. However, the indices may be considered as abstract, that do not refer to any particular system of coordinates, but conveniently denote the tensorial components \cite{Wal84b}. In the following we will use the indices in this sense. We do not distinguish between vectors and covectors. Einstein convention of repeated indices is applied.    

Secondary kinematic quantities are the relative deformation $\gamma^{ij} = \partial^iu^j-\psi^{ij}$, the microdeformation gradient $\mu^{ijk} = \partial^k\psi^{ij}$ and the velocity $v^i$. The velocity and strain fields are related by
\be
\dot \ep^{ij} = \partial^{(j} v^{i)}.
\label{kincon}\ee

The fundamental balance of momentum is:
\be\label{mombal}
\varrho \dot v^i - \partial^j t^{ij} = 0,
\ee
where $\varrho$ is the density, the dot denotes the substantial time derivative, the symmetric $t^{ij}$ is the stress and $\partial^i$ is the spatial gradient. The balance of internal energy follows as usual:
\be\label{intebal}
\varrho \dot e +\partial^iq^i = t^{ij}\partial^jv^i
\ee
where $e$ is the specific internal energy and $q^i$ is the current density of the internal energy, the heat flux. 
The evolution of the microdeformation is given as:
\be\label{Mevol}
\hat d^{ik} \ddot\psi^{jk} - \partial^{k}\mu^{ijk} = \sigma^{ij}.
\ee
Here $\hat d^{ik}$ is the tensor of microinertia, $\mu^{ijk}$ is the double stress and  $\sigma^{ij}$ is the relative stress. In the ideal, nondissipative theory of Mindlin this equation is derived from a variational principle and the three stresses are derived from a free energy. Generalized mechanics is a first order weakly nonlocal, gradient theory in the microdeformation therefore the free energy is a function of the strain, the relative deformation and the gradient of the microdeformation.  For our purposes it is  convenient to consider the free energy as a function of the microdeformation instead of the relative one. The two representations are not equivalent, because relative deformation depends also on the antisymmetric part of the displacement gradient. That seems to contradict to the usual formulations of material frame indifference \cite{TruNol65b}, therefore in the following we consider specific free energy as function of microdeformation instead of the relative one:  $w(\ep^{ij},\psi^{ij},\
partial^k\psi^{ij})$. Then the equilibrium stresses are expressed by the following derivatives of the free energy:
\be\label{dw}
t^{ij}_E = \frac{\partial w}{\partial \ep^{ij}}, \qquad
\sigma^{ij}_E = -\frac{\partial w}{\partial \psi^{ij}}, \qquad
\mu^{ijk}_E = \frac{\partial w}{\partial\left(\partial^k\psi^{ij}\right)}.
\ee

In case of isotropic materials a quadratic free energy has the following form:
\bea
&&w(\epsilon^{ij},\psi^{ij},\partial^k\psi^{ij}) = \nonumber\\
  &&\frac{\lambda}{2} (\epsilon^{ii})^2 + 
    \mu \epsilon^{ij}\epsilon^{ij} + 
    \frac{b_1}{2} (\psi^{ii})^2 + 
    \frac{b_2}{2} \psi^{ij}\psi^{ij} + 
    \frac{b_3}{2} \psi^{ij}\psi^{ji} + \nonumber\\
  &&g_1\psi^{ii}\ep^{jj}+
    g_2(\psi^{ij}+\psi^{ji})\ep^{ij}+\nonumber\\
  &&a_1 \partial^k\psi^{ii}\partial^j\psi^{kj} +  
    a_2 \partial^k\psi^{ii}\partial^j\psi^{jk} + 
    \frac{a_3}{2}\partial^k\psi^{ii}\partial^k\psi^{jj} +  
    a_4\partial^j\psi^{ij}\partial^k\psi^{ik} +  \nonumber\\
  &&a_5 \partial^j\psi^{ij}\partial^k\psi^{ki} + 
    \frac{a_6}{2} \partial^i\psi^{ij}\partial^k\psi^{jk} + 
    \frac{a_{7}}{2} \partial^k\psi^{ij}\partial^k\psi^{ij} +
    \frac{a_{8}}{2} \partial^k\psi^{ij}\partial^i\psi^{jk} + \nonumber\\
  &&\frac{a_{9}}{2} \partial^k\psi^{ij}\partial^j\psi^{ik} + 
    \frac{a_{10}}{2} \partial^k\psi^{ij}\partial^k\psi^{ji} + 
    \frac{a_{11}}{2} \partial^k\psi^{ij}\partial^i\psi^{kj}, 
\label{wfun}\eea
where $\lambda$ and $\mu$ are the Lam\'e coefficients, $b_1, b_2, b_3, g_1, g_2, a_1, ..., a_{11}$ are material parameters. Then, substituting the constitutive functions \re{dw} into the evolution equations \re{mombal}, \re{intebal} and \re{Mevol} we obtain a closed system. 

It is generally believed, that Mindlin's version of the microdeformation theory is nondissipative. Actually, without calculating the entropy production one cannot say anything about dissipation. The approach of Eringen and Suhubi is superior from this point of view, they calculate entropy production. On the other hand their dissipation -- by artificially introducing dissipative parts of the stress, relative stress and double stress -- is questionable. In classical continuum mechanics the entropy inequality is a conditional restriction, it is valid with the condition of the validity of evolution equations, the classical balances. In case of a second order evolution equation like \re{Mevol}, that condition is not compatible with energy dissipation formula of Eringen and Suhubi, eq.(5.13) in \cite{EriSuh64a}. Therefore in the following we suggest a simple alternative approach of the Mindlin--Eringen-Suhubi (MES) theory.

\section{Thermomechanics with dual internal variables}

Based on \cite{VanAta08a} one may propose a pure thermodynamic approach where the microdeformation $\psi^{ij}$ is an internal state variable and introduce also a second tensorial internal variable denoted by $\beta^{ij}$ \cite{BerEta11a}. Then we assume, that the entropy density is a function of the internal energy $e$, the strain $\ep^{ij}$, the internal variables $\psi^{ij}$, $\beta^{ij}$ and the gradient of the microdeformation $\partial^k\psi^{ij}$. Moreover, we do not require a variational principle or a balance of micromomentum, the evolution equations of the internal variables $\psi^{ij}$ and $\beta^{ij}$ are 
\be
\dot \psi^{ij} = f^{ij}, \qquad 
\dot \beta^{ij} = g^{ij}.
\label{evol}\ee

Here the right hand side of the equations are to be determined by the second law. We expect to obtain the MES theory as a nondissipative special case. Our only non straightforward assumption is the following, non classical form of the entropy current density:
\be\label{entrcurr}
J^i = \frac{q^i}{T} - \varrho\frac{\partial s}{\partial (\partial^k \psi^{ij})}f^{ij}.
\ee
Here $T$ is the temperature, defined by $\partial_es= 1/T$.

The entropy balance can be calculated as follows:
\bea
0 &\leq & \varrho\dot s(e,\ep^{ij},\psi^{ij},\partial^k\psi^{ij},\beta^{ij})) + \partial_i J^i = \nonumber\\
  &&\varrho \partial_e \dot e +
    \varrho \partial_{\ep^{ij}}s\dot \ep^{ij} +
    \varrho \partial_{\psi^{ij}}s\dot \psi^{ij}+
    \varrho \partial_{\partial^k\psi^{ij}}s(\partial^k\psi^{ij}\dot)+
    \partial_{\beta^{ij}}s\dot \beta^{ij}+ \nonumber\\
  &&\partial_i\left(\partial_es q^i -
      \varrho\partial_{\partial^i\psi^{jk}}s \dot\psi^{jk} \right) =\nonumber\\
  &&\frac{1}{T}\left(t^{ij} \partial^iv^j - \partial^iq^i \right) +
    \varrho \partial_{\ep^{ij}}s\partial^iv^j +
    \varrho \partial_{\psi^{ij}}s f^{ij}-
    \varrho \partial^k\left(\partial_{\partial^k\psi^{ij}}s\right)f^{ij}-\nonumber\\
  &&\varrho\partial_{\partial^k\psi^{ij}}s\partial^l\psi^{ij}\partial^kv^l +
    \varrho \partial_{\beta^{ij}}sg^{ij}+ 
    \partial^i\left(\frac{1}{T} q^i \right)  = \nonumber\\
  &&\frac{1}{T}\left(t^{ij} + \varrho T (\partial_{\ep^{ij}}s-
    \partial_{\partial^i\psi^{kl}}s\partial^j\psi^{kl})\right)\partial^iv^j +
    \varrho\left(\partial_{\psi^{ij}}s -
      \partial^k(\partial_{\partial^k\psi^{ij}}s) \right)f^{ij} +\nonumber\\
  &&\varrho\partial_{\beta^{ij}}s g^{ij}+  \left(q^i - 
      \varrho\partial_{\partial^i\psi^{jk}}s f^{jk}\right)\partial_i\frac{1}{T}      
\label{entrbal1}\eea
 
Here we have applied the kinematic condition \re{kincon} and the evolution equations of the momentum \re{mombal}, internal energy \re{intebal} and the internal variables \re{evol} as constraints. In small strain approximation the density can be considered constant. Remarkable are the nonclassical extensions of the mechanical and thermal contributions of the entropy production. Now we change the thermodynamic potentials and introduce the more traditional Helmholtz free energy representation. Then the transformations of partial derivatives are 
\bea
\left. \frac{\partial s}{\partial \epsilon^{ij}}\right|_e &=& 
    \left.-\frac{1}{T} \frac{\partial w}{\partial \epsilon^{ij}}\right|_T, \quad
  \left. \frac{\partial s}{\partial \psi^{ij}}\right|_e = 
    \left. -\frac{1}{T} \frac{\partial w}{\partial \psi^{ij}}\right|_T, \quad
  \left. \frac{\partial s}{\partial \beta^{ij}}\right|_e = 
    \left.-\frac{1}{T} \frac{\partial w}{\partial \beta^{ij}}\right|_T, \nonumber\\
\left. \frac{\partial s}{\partial \partial^k\psi^{ij}}\right|_e 
  &=& \left.-\frac{1}{T} \frac{\partial w}{\partial(\partial^k\psi^{ij})}\right|_T.
\eea\label{stow}

Substituting these formulas into \re{entrbal1} we obtain the following temperature multiplied form of the entropy production, that we call dissipation:
\bea
0 &\leq & TP_s = 
  \left(t^{ij}- \varrho (\partial_{\ep^{ij}}w-
      \partial_{\partial^i\psi^{kl}}w\partial^j\psi^{kl})\right)\partial^iv^j -\nonumber\\
&&    \varrho\left(\partial_{\psi^{ij}}w -
      \partial^k(\partial_{\partial^k\psi^{ij}}w) \right)f^{ij}-  \varrho\partial_{\beta^{ij}}w g^{ij}+  
    \left(q^i + 
      \varrho\partial_{\partial^i\psi^{jk}}w f^{jk}\right)T\partial_i\frac{1}{T}     
\label{diss}\eea

This is a quadratic form of terms that are products of the constitutive functions $t^{ij}$, $f^{ij}$, $g^{ij}$ and given functions of the constitutive state space. The last two constitutive functions determine the evolution equations. In this case we can apply a linear approximation of the constitutive functions in order to solve the inequality. It is legitimate to introduce thermodynamic fluxes and forces in the sense of non-equilibrium thermodynamics:
\begin{center}
\begin{tabular}{c|c|c|c|c}
       & Thermal & Mechanical & Internal 1 & Internal 2 \\ \hline
Fluxes & $\hat q^i =  q^i + $ & 
    $t_v^{ij} = t^{ij} - \varrho \partial_{\epsilon^{ij}} w +$ & 
    $\varrho f^{ij}$ &
    $\varrho g^{ij}$\\ 
       & $\varrho f^{jk} \partial_{\partial^i\psi^{jk}}w$ & 
       $\varrho\partial_{\partial^i\psi^{kl}}w\partial^j\psi^{kl}$ & & \\ \hline
Forces &$\partial^i \ln T$ &
    $\partial^{(i}v^{j)}$ &
    $X^{ij} = $ &
    $Y^{ij} = -\partial_{\psi^{ij}}w +$\\
    & & & $ \partial^k \partial_{\partial^k\psi^{ij}}w $& 
    -$\partial_{\beta^{ij}}w$ 
\end{tabular}\\
\vskip .21cm
{Table 1. Thermodynamic fluxes and forces of generalized mechanics}\end{center}

Here $t^{ij}_v$ denotes the viscous stress. In case of isotropic materials the linear relations between the thermodynamic forces and fluxes are the following:
\bea \label{lOns1}
\hat{q}^i &=& 
  \lambda \partial_i \ln T \\ \label{lOns2}
  t_v^{ij} &=&
    L_{11}^{ijkl} \partial^{(l}v^{k)} + 
    L_{12}^{ijkl}   X^{kl} + 
    L_{13}^{ijkl}   Y^{kl} \\ \label{lOns3}
\varrho f^{ij} &=& 
    L_{21}^{ijkl} \partial^{(l}v^{k)} + 
    L_{22}^{ijkl}   X^{kl} + 
    L_{23}^{ijkl}   Y^{kl} \\ \label{lOns4}
\varrho g^{ij} &=& 
    L_{31}^{ijkl} \partial^{(l}v^{k)} + 
    L_{32}^{ijkl}   X^{kl} + 
    L_{33}^{ijkl}   Y^{kl},
\eea
where for all $I,J= 1,2,3$ the conductivity coefficents $L_{IJ}$ are
\be \label{liso1}
 L_{IJ}^{ijkl} = 
  s_{IJ} \delta^{i\langle k}\delta^{jl\rangle} + 
  a_{IJ} \delta^{i[k}\delta^{jl]} + 
  l_{IJ} \delta^{ij}\delta^{kl},  
\ee
where the braces $\langle\ \rangle$ denote the traceless symmetric part of the corresponding tensor in the related indices 
$\delta^{i\langle k}\delta^{jl\rangle} = (\delta^{ik}\delta^{jl}+\delta^{il}\delta^{jk})/2 -\delta^{ij}\delta^{kl}/3$  
and the rectangular parenthesis $[\ ]$ denotes the antisymmetric part as  
$\delta^{i[k}\delta^{jl]} = (\delta^{ik}\delta^{jl}-\delta^{il}\delta^{jk})/2$. 
This kind of decomposition is instructive because symmetric, antisymmetric and spherical second order tensors are mutually orthogonal in the ``double dot'' product of second order tensors, when taking the trace of their product. Therefore the constitutive equations \re{lOns2}-\re{lOns4} can be decomposed into three parts: the five component traceless symmetric, three component antisymmetric and one component spherical parts are independent. 
Moreover, the vectorial and tensorial thermodynamic interactions are independent, too. 


The evolution equations of generalized mechanics are obtained when the internal variable $\beta^{ij}$ is formally a generalized momentum of the internal variable $\psi^{ij}$. One can see that applying two conditions. 

\begin{enumerate}
 \item \textit{Quadratic free energy.} The free energy of Mindlin \re{wfun} must be supplemented by quadratic isotropic $\beta^{ij}$ dependent terms:
\be
  w(\epsilon^{ij},\psi^{ij},\partial^k\psi^{ij},\beta^{ij}) = 
  w_{M}(\epsilon^{ij},\psi^{ij},\partial^k\psi^{ij}) +
 \frac{1}{2} \beta^{ij}B^{ijkl}\beta^{kl}.
\label{wfunadd}\ee
and $B^{ijkl} =b_s \delta^{i\langle k}\delta^{jl\rangle} + 
  b_a \delta^{i[k}\delta^{jl]} + 
  b_l \delta^{ij}\delta^{kl}$.  
 \item \textit{Ideality}. This condition switches off some dissipative couplings. For example for the  conductivity coefficients we prescribe that $L^{ijkl}_{13} = L^{ijkl}_{31}=L^{ijkl}_{33}=0^{ijkl}$ and require antisymmetric coupling of the evolution equations of the two internal variables by $L^{ijkl}_{32}=-L^{ijkl}_{23}=L^{ijkl}$.  
\end{enumerate}

Then taking the time derivative of \re{lOns3} and substituting $\dot\beta^{ij}$ from \re{lOns4} results in the following second order evolution equation of the microdeformation $\psi^{ij}$: 
\be
\varrho I^{ijkl} \ddot \psi^{kl} -\partial^k\mu^{ijk}_E  = \sigma^{ij}_E +  
  I^{ijkl}L_{21}^{klmn} \dot\ep^{mn} +
  I^{ijkl}L_{22}^{klmn}(\dot\sigma^{mn}_E + \dot\partial^k\mu^{mnk}_E)
\label{DissMind}\ee

Here $I^{ijkl} = L^{ijmn}B^{mnop}L^{opkl}$ is the microdensity tensor. 

The evolution equations \re{kincon}, \re{mombal}, \re{intebal} and \re{DissMind}, together with the dynamic constitutive relations \re{lOns1}-\re{lOns4} and static constitutive relations \re{wfun}, \re{stow} and \re{wfunadd} form a dissipative generalization of the MES theory except the kinematic interpretation of the internal variables. The evolution equations of the original nondissipative Mindlin-Eringen-Suhubi theory is recovered if $\lambda=0$ and $L_{IJ}^{ijkl}=0$ for $I,J=1,2,3$ except $L^{ijkl}_{23}=-L^{ijkl}_{32}=L^{ijkl} \neq 0$.

\section{Generic stability in one space dimension}

In this section we investigate the conditions of generic stability in case of a single space dimension, when the fields depend on the time $\tau$, and a single spatial coordinate, denoted by $x$. The extension to three space dimensions is straightforward. 

The one dimensional forms of the relation of the strain and the velocity \re{kincon}, the evolution equations of momentum \re{mombal}, internal energy \re{intebal} are the following
\bea
\dot\ep -\partial_x v &=& 0. \label{kincon1}\\
\varrho \dot v - \partial_x t &=& 0, \label{mombal1}\\
\varrho \dot e + \partial_x q &=& t \partial_x v, \label{intebal1}
\eea

Here Cartesian coordinates are introduced and the first, $x$ component of vectors and tensors is used in the reduction: $v(\tau,x) = v^x(\tau,x)$, $q(\tau,x)=q^x(\tau,x)$ are the $x$ components of the velocity and heat current fields, $t(\tau,x)=t^{xx}(\tau,x)$, $\ep(\tau,x)=\ep^{xx}(\tau,x)$ are the $11$ components of the stress and the strain fields. $\partial_x$ is the derivative by $x$ and the dot denotes sustantial time derivative, For example $\dot e = \partial_\tau e + v\partial_x e$ is the substantial time derivative of the specific internal energy and $\partial_\tau$ is the partial time derivative. 
The evolution equations \re{evol}  of the internal variables $\psi(\tau,x)=\psi^{xx}(\tau,x)$ and $\beta(\tau,x)=\beta^{xx}(\tau,x)$ are reduced to 
\bea
 \dot \psi  = f, \label{evol11}\\ \qquad \dot\beta = g.
\label{evol21}\eea

Here $f=f^{xx}$ and $g=g^{xx}$. The quadratic, isotropic free energy \re{wfunadd} in one space dimension is
\be
  w(\ep,\psi,\partial_x \psi,\beta) = 
  \frac{\hat E}{2} \ep^2 + \frac{\hat b}{2} \psi^2 + \hat c\ \ep\psi+\frac{\hat B}{2} \beta^2 + \frac{\hat a}{2} (\partial_x\psi)^2.  
\label{wfun1}\ee

The convexity of the free energy requires, that the  $\hat E,\hat b,\hat B,\hat a$ coefficients are positive and $\hat E\hat b-{\hat c}^2 \geq0$.
The dissipation \re{diss} is obtained as
\bea
 0 &\leq& 
  -\partial_x (\ln T) (q+\varrho f \partial_{\partial_x\psi}w)+
  (t-\varrho(\partial_\ep w-\partial_{\partial_x\psi}w \partial_x w))\partial_xv - 
  \nonumber\\
  && \left(\partial_\psi w- \partial_x(\partial_{\partial_x \psi}w)\right)\varrho f - 
  \left(\partial_\beta w\right)\varrho g.
\label{edis1}\eea

The consequent linear conductivity relations are 
\bea \label{lOns11}
\hat{q} &= q +\varrho f \partial_{\partial_x\psi}w &= 
  -\lambda \partial_x \ln T \\ \label{lOns21}
  t_v  &=  t - (E\ep+c\psi)+ a(\partial_x\psi)^2 &=
    l_{11}\partial_xv + l_{12}\varrho X + l_{13}\varrho Y \\ \label{lOns31}
 \dot{\psi} &=  f &= 
    l_{21}\partial_xv + l_{22}\varrho X + l_{23}\varrho Y  \\ \label{lOns41}
 \dot{\beta} &= g &= 
    l_{31} \partial_xv + l_{32}\varrho X + l_{33}\varrho Y .
\eea
where $\varrho X=-\varrho\partial_\psi w+ \varrho\partial_x(\partial_{\partial_x \psi}w) = -\tilde b \psi - c \ep + a\partial_{xx}\psi$, and $\varrho Y=-\varrho\partial_\beta w = -B \beta$ and $E = \varrho \hat E$, $\tilde b = \varrho \hat b$, $c = \varrho \hat c$, $B = \varrho \hat B$ and $a = \varrho \hat a$. The inequality of second law \re{edis1} requires, that $\lambda\geq 0$ and the symmetric part of the matrix $l_{IJ}$, $I,J=1,2,3$ is positive definite. Here the known transport coefficients in $l_{IJ}$ are $l_{11}$, the coefficient of viscosity, and $L = (l_{23}-l_{32})/2$, the antisymmetric coupling term of the dual variables, from the microdensity $I=L^2B$.

\subsection{Homogeneous equilibrium}

We consider the kinematic condition \re{kincon1}, the evolution equations \re{mombal1}-\re{evol21} and the constitutive functions in \re{lOns11}-\re{lOns41}, defining $\hat q$, $t$, $f$ and $g$. It is convenient to introduce the stress $t$ as an independent variable in addition to $e,v,\ep,\psi$ and $\beta$. Let us denote the homogeneous solution of these equations by $e_0,v_0,\ep_0,t_0,\psi_0,\beta_0$. The space derivative of these fields vanishes, they are functions only of time. Therefore from \re{kincon} follows $\ep_0=const.$ and we choose the homogeneous equilibrium  strain zero. From the momentum balance \re{mombal} we obtain, that $v_0=const.$, and from the energy balance \re{intebal} follows that $e_0=const.$. The homogeneous equilibrium values of the internal variables are determined by the differential equations:
\be
  \dot \psi_0  = -l_{22} \tilde b\psi_0 - l_{23}B\beta_0, \qquad
  \dot \beta_0 = -l_{32} \tilde b\psi_0 - l_{33}B\beta_0
\label{rela}\ee

The homogeneous equilibrium stress is  $t_0 = (c-l_{12} \tilde b)\psi_0  - l_{13}\beta_0$. We can see, that the coefficents of the above equations are nonnegative, therefore $\psi_f = 0$, $\beta_f =0$, and $t_f=0$ are asymptotic equilibrium values of the relaxation equations \re{rela}. 

\subsection{Linearization}

In the following we assume small perturbation around the homogeneous equilibrium solutions, therefore  the fields $\xi=(e,v,\ep,t,\psi,\beta)$ have the form $\xi(\tau,x) = \xi_0(\tau)+\delta\xi(\tau,x)$. Then the linearization of the equations \re{kincon1}-\re{evol21} and \re{lOns11}-\re{lOns41} result in:
\bea
\partial_\tau \delta\ep &=&\partial_x \delta v , \label{linkincon}\\
\varrho \partial_\tau \delta v &=&\partial_x \delta t, \label{linmombal}\\
\varrho \partial_\tau \delta e &=&\lambda \frac{(\partial_e T)_0}{T^2} \partial_{xx} \delta e, \label{linintebal}\\
\delta t &=&  E\delta\ep+ c\delta \psi + l_{11} \partial_x \delta v - 
  \varrho l_{12} (\tilde b \delta \psi+c \delta \ep - a\partial_{xx} \delta \psi)-
  \varrho l_{13} B\delta\beta, \label{stresscal}\\
\partial_\tau\delta\psi  &=&  l_{21} \partial_x \delta v - 
  \varrho l_{22} (\tilde b \delta \psi+c \delta \ep - a\partial_{xx} \delta \psi)-
  \varrho l_{23} B\delta\beta, \label{linevol1}\\  
\partial_\tau\delta\beta  &=&  l_{31} \partial_x \delta v - 
  \varrho l_{32} (\tilde b \delta \psi+c \delta \ep - a\partial_{xx} \delta \psi)-
  \varrho l_{23} B\delta\beta, \label{linevol2}  
\eea

Here $T_0$ is the temperature in the equilibrium and we have assumed that the temperature is the function of the internal energy and independent of the other state variables. Then \re{linintebal} is independent of the other equations.  Looking for conditions of stability we introduce exponential wave perturbations of the following form: $\delta\xi(t,x) = \delta_0 e^{\Gamma t +ikx}$. This exponential wave solution of the linear system of equations \re{linkincon}-\re{linevol2} is damped if the real part of $\Gamma$ in the solutions is negative. The linearized balance of internal energy \re{linintebal} gives:
\be
\varrho\Gamma  \delta e= -\lambda \frac{(\partial_e T)_0}{T^2} k^2 \delta e.
\ee
Therefore $\Gamma =  -\lambda \frac{(\partial_e T)_0}{T_0^2\varrho} k^2$ is negative because the Fourier heat conduction coefficient $\lambda_F/T^2$ is positive and the specific heat $\partial_eT$ is positive to ensure nonnegative entropy produstion and concave specific entropy function respectively.

The remaining system of equations \re{linkincon}, \re{linmombal}, \re{stresscal}-\re{linevol2} is transformed to the following linear matrix form:
\be\label{linmatr}
\begin{pmatrix}
\Gamma       & -i k          & 0   & 0                               & 0         \\ 
0            & \varrho\Gamma & -ik & 0                               & 0         \\
-E+ l_{12} c & -l_{11}ik     & 1   & -c+l_{12}(\tilde b+a k^2)       & -l_{13} B \\
l_{22} c     & -l_{21}ik     & 0   & \Gamma + l_{22}(\tilde b+a k^2) & l_{23} B  \\
l_{32} c     & -l_{31}ik     & 0   & l_{32}(\tilde b+a k^2)          & \Gamma+l_{33}B
\end{pmatrix}
\begin{pmatrix}
\delta\ep \\
\delta v \\
\delta t\\
\delta\psi \\
\delta\beta
\end{pmatrix} = 0.
\ee

Exponentially growing wave solutions of \re{linmatr} emerge whenever $\Gamma$ and $k$ satisfy the dispersion relation obtained from the determinant of the matrix above where $\Gamma$ has positive real parts. In order to simplify the calculations we assume that $l_{13} = l_{31} = 0$ and $l_{12} = l_{21}$. In this case we preserve the viscosity $l_{11}$, the other dissipative terms in \re{DissMind} $l_{22}$ and $l_{12}$, and keeping also $l_{33}$ we have a term that is beyond the MES framework.

The determinant is a fourth order polinomial in $\Gamma$:
\bea
  \varrho \Gamma^4 + 
  \left[\varrho( b l_{22}+Bl_{33})+ k^2 l_{11}\right]\Gamma^3+\nonumber\\
  \left[\varrho b B\Delta_{23} + k^2 (E+ b\Delta_{12} + Bl_{33}l_{22})\right]\Gamma^2 +\nonumber\\
  k^2 \left[l_{22}\delta + b B\Delta + BEl_{33}\right]\Gamma+
  k^2 B \delta\Delta_{23} =0.
\label{sdet}\eea

Here $\Delta_{23} = l_{22} l_{33}- l_{23}l_{32}>0$, $\Delta_{12} = l_{22}l_{11}- l_{12}l_{12}>0$, $\Delta = l_{11}l_{22}l_{33}- l_{11}l_{23}l_{32} -l_{33}l_{12}l_{21}>0$ are positive, because the symmetric part of $l_{IJ}$ is positive definite. We have introduced $b= \tilde b+ak^2$ and $\delta =E\tilde b-c^2$ is nonnegative because of the convex free energy. Therefore the coefficients of the above polinomial are nonnegative. According to the Rout-Hurwitz criteria the real parts of a fourth order polinomial $a_4x^4+a_3x^3+a_2x^2+a_1x+a_0 =0$ are negative, whenever the coefficients of the polinomial are positive and \cite{KorKor00b}
\be
C_1 = a_3a_2 - a_4a_1 \geq 0, \qquad 
C_2 = (a_3a_2 - a_4a_1)a_1- a_0a_3^2 \geq 0
\ee
In our case for \re{sdet} the first condition gives 
\bea
C_1 = &k^4&\!\!\!\!\! l_{11} (\mu+Bl_{11}l_{33}+  b\Delta_{12})+ \nonumber\\
&k^2&\!\!\!\!\! \varrho (  b^2l_{22}\Delta_{12}+B^2l_{11}l_{33}l_{33}+l_{22}(g^2 + 2 bBl_{11}l_{33}))+ \nonumber\\
&&\!\!\!\!\! bB\varrho (  b l_{22}+Bl_{33})\Delta_{23}\geq 0
\label{c1}\eea
One can see, that all terms are positive in the above expression.

The second condition is more complicated: 
\bea
  C_2 &=&\nonumber \\
  &k^6&\!\!\!\!\! l_{11}\left[-B\delta l_{11} \Delta_{23} + 
    (\mu\! +\! Bl_{11} l_{33} + b\Delta_{12})(l_{22}\delta + b Bl_{11}\Delta_{23} + 
       B l_{33}(\mu\!\! -\! bl_{12}^2))\right]\! \!+ \!\!\nonumber\\
  &k^4&\!\!\!\!\! \varrho \left[-2 B\delta l_{11}(  bl_{22} + Bl_{33})\Delta_{23} + 
    (  b^2 l_{22} \Delta_{23} + B^2 l_{11}l_{33}^2+ 
    l_{22} B(B + 2  bl_{11}l_{33}))\right.\!\!\!\times \nonumber\\
   && \left.\times (-l_{22}\delta + 
         b Bl_{11}\Delta_{23} + Bl_{33}(\mu -   b l_{12}^2))\right] +\nonumber\\
  & k^2 & \!\!\!\!\! B^2 \varrho^2 (bl_{22} + Bl_{33})\Delta_{23} (l_{33} B^2 + b^2 \Delta)
\label{c2}\eea

Some transformations may be usefull here. The coefficient of the $k^6$ term can be written as
\bea
  \frac{\delta^2}{b^2}(bl_{22}+Bl_{33})+
  \frac{\delta}{b^2}\left[b^2(bl_{22}+2Bl_{33})\Delta_{12} + 
    B^2(Bl_{33}+2bl_{22})+  bB^2l_{11}l_{33}^2\right]+ \nonumber\\
  \frac{1}{b^2}B(B^2 + bB(\Delta_{12}+l_{11}l_{33}))(B^2l_{33}+b^2\Delta),
\label{k6}\eea
and that is clearly positive. The coefficient of the $k^4$ terms follows as
\bea
  \frac{B}{b}(b^2l_{22}\Delta_{12} + B^2l_{11}l_{33}^2+
    l_{22}(B^2+2 b Bl_{11}l_{33})) + \nonumber \\
  \frac{\delta}{b}(Bl_{33}+bl_{22})\left[b^2 l_{22}\Delta_{12} + l_{22} B^2+ 
    Bl_{11}(Bl_{33}^2+2bl_{23}l_{32}\right]
\label{k4}\eea
This is positive, too, except of the last parenthesis in the last term, where $l_{32}l_{23}$ may be negative. It is reasonable to consider that term as a second order polinom of $b$, which is positive, if the discriminant is negative. This condition is:
\be
 Disc= l_{22}(l_{33})^2\Delta_{12} - l_{11}(l_{23}l_{32})^2 <0.
\ee
This inequality can be rewritten in the following form:
\be
  l_{33}\Delta_{23}\Delta_{12} + l^{2}\tilde\Delta + l_{11}L^2(2l^2-L^2) >0
\label{ficon}\ee

Where we have introduced the $l=(l_{23}+l_{32})/2$ and $L=(l_{23}-l_{32})/2$ for the symmetric and antisymmetric parts of the couplings of $\psi$ and $\beta$ evolution. $\tilde\Delta$ is the determinant of the symmetric part of $l_{IJ}$
\be
  \tilde\Delta = l_{11}l_{22}l_{33}-l_{33}l_{12}l_{12}-l_{11}l^2
\ee

The final condition \re{ficon} cannot be violated if $2l^2>L^2$, however the last term may dominate the previous ones for sufficiently large $L$. This can happen only if also $l_{33}$ or $l_{22}$ is not zero, otherwise the second term in \re{k4} is zero. Therefore dissipation is necessary for the violation of generic stability conditions.

\section{Discussion}

We have introduced a reasonably simple dissipative version of the microdeformation theory of Mindlin-Eringen-Suhubi. The generalization is based on thermodynamic principles extending the basic state space of continuum mechanics by dual weakly nonlocal internal variables. The most important distinctive property of the generalization is the nonclassical form of the entropy current density when the entropy density depends on the gradient of the state variables. There are two remarkable predictions of the new approach. The first one is the extra term in the thermodynamic force of heat conduction beyond the temperature gradient \cite{BerEta11a}. The second one is the modification of the nondissipative part of the stress. The second contribution is related the question of material time derivatives for the internal variables. A more detailed and more rigorous analysis of the thermodynamic approach to generalized mechanics based on the exploitation of the entropy inequality in case of second order weakly nonlocal 
constitutive state spaces in first Piola-Kirchhoff framework is given in \cite{VanEta13m}. 

In a simple but reasonably representative example in one space dimension we have obtained the conditions of generic stability of the theory. We have shown that in some particular cases the thermodynamic conditions alone do not ensure generic statibility. 

The stability of homogeneous equilibrium in case of small perturbations is a basic physical expectation in any dissipative continua, therefore we suggest generic stability as a benchmark for testing the viability of dissipative continuum theories both in classical and relativistic spacetimes.

\section{Acknowledgement}   
The work was supported by the grant Otka K81161. The authors thank to Tam\'as F\"ul\"op, Csaba Asszonyi, Tam\'as Matolcsi and Arkadi Berezovski for valuable discussions.

\bibliographystyle{unsrt}

\end{document}